\documentstyle[12pt,aaspp4,epsfig]{article}

\slugcomment{\it To appear in the Astrophysical Journal}

\begin{document}

\title{The Mass and Mass-to-Light Ratio for Clusters of Galaxies at Large
Radii}

\author{Henrik Vedel}
\affil{Theoretical Astrophysics Center, Juliane Maries Vej 30,
\\ DK-2100 Copenhagen, Denmark}

\author{F.D.A. Hartwick}
\affil{Dept.\ of Physics \& Astronomy,
P.O. Box 3055, University of Victoria \\
Victoria, B.C. V8W 3P6, Canada}

\centerline{e-mail: vedel@tac.dk and hartwick@uvastro.phys.uvic.ca}

\begin{abstract}

We construct models describing the velocity field in the infall regions of
clusters of galaxies. In all models the velocity field is the
superposition of a radial systematic component which is assumed
spherically symmetric, and a ``noise" component of random nature. The
latter accounts for the combined effects of small-scale substructure and
observational errors. The effect of the noise term is to smear out the
caustic envelopes searched for by previous groups, resulting in models
which resemble the observations quite well. When the systematic component
is known, the infall velocity and the mass profile of the infall region of
the clusters can be determined.

Given a particular model, it is possible to calculate the distribution
function, $f(cz,\theta,m)$, at all points in the observable
$(cz,\theta,m)$-phase-space outside the virialized region. It is demonstrated,
on simulated data, that the use of a maximum-likelihood estimator enables
identification of the correct model, even at realistic noise levels. To
minimize systematic effects due to deviations from spherical infall,
observations from a number of clusters should be used when applying the
method to a real dataset. Combining data from different clusters is a
straight-forward procedure.

In the models the mass-to-light ratio is allowed to vary freely with
radius. Furthermore, the models do not require any knowledge of the virial
region. Rather, such knowledge can be used to test the models ability to
produce realistic results by comparing the mass estimated by the models at
small radii to the mass estimates obtained for the inner region using the
virial theorem.

As a corollary we show how distance information of the quality obtained
using the Tully-Fisher relation enhances the likelihood signal. Such
distance information might be used in the future to either strengthen the 
results for
this type of model or allow more advanced models to be used (e.g. models
breaking the assumption of spherical symmetry).

The method relies on observations of redshifts of galaxies at large
angular separations from the centers of the clusters. Currently such data
exist only for Coma, to which we apply the method as an example. The
estimated mass at $r=2.5h^{-1}Mpc$ falls in the range
$1.1-2.4\times10^{15}h^{-1}M_\odot$, depending on the model and the level
of the noise term. We find $r_{turn}$ to be of the order $10 h^{-1}Mpc$, a
result which is rather robust against changes in the model and subdivision
of the data in various ways. We find $0.6 <
\frac{\Omega_0^{0.6}}{b^{0.75}} < 0.8$, where $b$ is the bias parameter,
when using the luminosity profile adopted in the models. Interpreting this
result one must recall that the luminosity function of Coma is not well
known in the infall region, and that ideally the method should be used on
several clusters simultaneously to minimize the effects of deviations from
the spherical model. 

The method provides a promising way of measuring the infall velocities in
the outer regions of clusters. However, a robust determination of cluster
properties must await future redshift observations of galaxies in the outer
regions of a number of clusters.

\end{abstract}

\keywords{galaxies: clustering --- clusters: mass, mass-to-light ratios}

\clearpage

\section{INTRODUCTION}

Current estimates of $\Omega_0$ based on the mass-to-light ratio found for
the virialized central regions of clusters of galaxies favour a low value
(Carlberg et al.\ 1997a, David, Jones, \& Forman 1995, Tyson
\& Fisher 1995, Squires et al.\ 1996, Smail et al.\ 1997, Fisher \& Tyson
1997), of the order $\Omega_0 \approx 0.25$. Estimates from much
larger scales often favour higher
values, including $\Omega_0=1$, but the uncertainties are in general
rather large (see Dekel, Burstein, \& White 1996 for a recent review).

If the global value of
M/L is significantly higher than the value found for the central parts of
clusters it would not be unnatural for the change to manifest itself 
in the infall region of the clusters. 
If, for example, the difference is due
to environmental effects resulting in the baryons being "lit up" depending
on the properties of the surroundings, one can argue that
the infalling galaxies formed well away
from the clusters and have yet to be strongly affected by cluster imposed
tidal fields, hot cluster gas, etc. 
Furthermore numerical simulations indicate that the extent of the
mass density profiles of clusters is related to $\Omega_0$, more 
extensive profiles corresponding to higher values of $\Omega_0$ due
to ongoing infall (Crone, Evrard, \&
Richstone 1994). The steepness is also related to the precise form of the 
initial spectrum of density fluctuations, however. 

Obtaining mass profiles of the outer regions of clusters is therefore
highly relevant.

Three main measures are being used for determination of the mass-to-light
ratio of clusters: the velocity dispersion, the temperature of the hot
X-ray emitting gas in the clusters, and the distortion of the shapes of
background galaxies. The first two rely on the assumption of virial
equilibrium and are consequently not the ideal tools for a determination
of the mass at significantly larger radii than the virial radius, at least
not without some modification of the approach governed by insight from
N-body simulations. Observation of weak gravitational lensing using new
8--10-m telescopes and large array CCDs is a very promising technique
which potentially can probe the cluster potentials far into the infall
region. However, at present the mass estimates found from the
weak lensing method generally apply to the inner regions.

In this paper we shall develop a second method; we attempt to measure the
systematic part of the velocity field of the galaxies in the infall
region. 

As the peculiar velocities, relative to those expected for a pure Hubble
flow, will be larger near a cluster than in the general field one might
expect it to be a less difficult task to map the peculiar velocity field
of the galaxies surrounding a cluster rather than the velocity field on
scales on which linear theory applies. However, even though the infall signal
should be rather strong , projection effects and our inability to 
measure distances of galaxies precisely makes the determination difficult.

Kaiser (1987) applied the spherical infall model (e.g. Lilje \& Lahav
1991) to an (adopted) spherical cluster density profile and found that it
manifest itself as a characteristic diamond shaped distribution in the 
distribution of galaxies in the $(cz, \theta)$-plane due to caustics.
(here $cz$ is the
redshift and $\theta$ is the angular 
separation between a given galaxy and the cluster center).  

Reg{\"o}s \& Geller (1989, RG) extended the work of Kaiser and showed that
the shapes and sizes of the caustics are a characteristic of the
mass enclosed. RG assumed light to trace mass, in which case $\Omega_0$
becomes the only free parameter of the model, as the observed luminosity
profile can be deconvolved and used to deduce the relative mass
over-density as a function of radius.
Fitting their models by eye to observations of the Coma-cluster (and a few
other clusters) RG were not able to put new constraints on $\Omega_0$,
however. The data available to them were limited, and what they had appeared
to be more noisy than predicted by their models. The density gradients
across the caustics seemed to be weaker than predicted by their model.

Van Haarlem et al.\ (1993, vH) adopted the basic assumptions of RG and
refined their work on the Coma cluster in three ways: Firstly, they
introduced an objective measure of the change in density between the
region just inside and just outside the position of the caustics predicted
by a given model, allowing a more rigorous fitting procedure. Using Monte
Carlo simulated artificial clusters the new measure was shown to work
well. Secondly they extended the number of observations available by
performing a small redshift survey around the Coma cluster for $\theta
[1^{\circ}; 3^{\circ}]$, adding 98 new redshifts to the Coma sample.
Thirdly, using scans of Schmidt plates, they determined the luminosity
profile of Coma as well as is currently possible (assuming spherical
symmetry).

Despite this the result was still inconclusive, which they
noted was due to the caustics being smeared out in the observational
sample, whereas the model caustics were very sharp. vH attributed the
smearing observed in real clusters to a non-spherical mass distribution,
and they demonstrated that an ellipsoidal mass distribution will produce a
distinct smearing of the infall in the $(cz, \theta)$-plane.
Later work (e.g. van Haarlem \& van de Weygaert 1993),  
show the non spherical distribution of the infalling material (infall along 
filaments), and the event-like nature of part of the infall, to be important
factors, however.
 
In this paper we show that keeping the spherical approximation for the infall,
but adding noise to the models, noise which is known to exist in
all clusters, and which
represents the combined effects of the velocities induced by small
scale density fluctuations and observational errors, can be responsible for 
much of the smearing of the caustics. We are able to produce artificial 
clusters with our models which appear quite like real clusters in the 
$(cz,\theta)$-plane.

The organisation of the paper is as follows: Section 2 describes the
models, the maximum likelihood procedure used to fit models to data, and
tests of the procedure on simulated data. In Section 3 the models are
applied to the Coma cluster as an example. Section 4 contains a discussion
of our findings and assumptions, and Section 5 the conclusions.

\section{MODELS AND TESTS OF MODELS}

We will construct models which allow us to determine the infall velocity
profile and mass profile of clusters of galaxies beyond the virial region,
based upon observations of the position on the sky, the line-of-sight
velocity, and the magnitudes of individual galaxies in and around the
clusters.

We shall attempt to perform this task by first constructing a series of
simple models of the velocity field of the region beyond the distance of
shell crossing, with each model having only a few free parameters to be
determined. Secondly, the distribution function, $f(cz,\theta,m)$, which
expresses the number density of galaxies expected at any given point in
the observed region of phase-space is calculated for each model. Finally,
the best-fitting parameters are found using maximum-likelihood techniques.

With the model-parameters in hand, estimates of the basic properties of the
outer regions of a cluster, such as the mass inside a given radius, the
mass-to-light ratio (M/L), and the turn-around radius can be estimated,
and eventual conclusions about $\Omega_0$ can be drawn.

In all of the following we define the ``infall region" as the region of a
cluster which is {\it outside} the part seriously effected by processes
associated with virialization and shell crossing and {\it inside} the
volume outside which neighbouring clusters have a significant impact on the
peculiar velocities of the galaxies relative to those induced by the
cluster itself. We shall apply our models only to the infall region, as
here defined. It contains galaxies which are falling towards the cluster
for the first time as well as more distant galaxies which are still moving
away from the cluster.

We shall assume that the velocity field in the infall region can be
adequately described by a systematic component which is the one we want to
determine plus a `noise' component of random nature. The latter allows for
the effects of substructure, non-radial motions, and observational errors.

Following RG, we will let the systematic component of the velocity field
be given by a spherically symmetric radial flow. Many clusters show
evidence for deviations from spherical flow, but combining data from
a number of clusters we expect the effects of non-sphericity to be minimized
out, resulting in a spherical cluster while adding some extra noise to the
infall signal.

\subsection{Infall Velocity Versus Density in the Spherical Approximation}

The spherical approximation is based on the constraint that one is dealing
with a spherically symmetric density fluctuation and the corresponding
purely radial velocity fluctuation in an otherwise perfectly homogeneous
and isotropic background universe. In the limit of no shell crossings
(i.e. there is no mass transport through the shell over the time span
considered) the evolution of the radius of a shell and the current time
are related through the following equations:
\begin{equation}
\frac{r}{r_i} =\frac{a}{|b|}f(\eta) ~~~{\rm and}~~~
tH_0=\frac{a}{|b|^{3/2}}g(\eta),
\end{equation}
where:
\begin{equation}
f(\eta) = \left\{
\begin{array}{ll}
1-\cos\eta, & b > 0 \\ \cosh\eta -1,
& b < 0
\end{array}
\right.
~~{\rm and}~~
g(\eta) = \left\{
\begin{array}{ll}
\eta-\sin\eta, & b > 0 \\ \sinh\eta-\eta, & b < 0
\end{array}
\right. ,
\end{equation}
where $a, b, r_i$ are constants related to the initial properties of the
shell and to the type of cosmology (e.g. Peebles 1980). Throughout this
paper we use $100hMpc$ km$^{-1}$ s$^{-1}$ for the value of the Hubble
constant H$_0$. 

It follows that the ratio of today's velocity, $v(r)$, of a given shell at
distance $r$ from the center of the perturbation to the velocity
difference expected over the same distance in a situation with pure Hubble
flow is:
\begin{equation}
\frac{v}{H_0r}=\frac{1}{t_0H_0}\frac{g(\eta)f'(\eta)}{f^2(\eta)},
\end{equation}
whereas the ratio of today's mean density inside the same shell,
$\overline{\rho}(r)$, to that of the background universe is given by:
\begin{equation}
\frac{\overline{\rho}}{\rho_0}=\frac{2}{(t_0H_0)^2\Omega_0}
\frac{g^2(\eta)}{f^3(\eta)},
\end{equation}
(e.g. Silk 1977).
Remembering that:
\begin{equation}
 t_0H_0 = \left\{ \begin{array}{ll}
\frac{1}{1-\Omega_0}-\frac{\Omega_0}{2(1-\Omega_0)^{2/3}}
\cosh(\frac{2}{\Omega_0}-1), & \Omega_0 < 1 \\ \frac{2}{3}, &
\Omega_0=1, \end{array} \right.
\end{equation}
and
\begin{equation}
\rho_0=\frac{3 \Omega_0 H_0^2}{8 \pi G}
\end{equation}
it is possible to determine the mass profile in the infall region of a
cluster if one can somehow measure its velocity profile. When combined
with the luminosity profile of the cluster, M/L can be determined.

Unfortunately, it is currently not possible to measure the line-of-sight
velocity as a function of distance directly, from which $v(r)$ could be
obtained. Distance estimates for galaxies are far from being precise
enough. One is forced to work primarily with angular position plus
redshift data, and consequently $v(r)$ must be obtained in a much more
indirect way.

In this work we do so by constructing a model which predicts the
distribution of $cz$ as a function of $\theta$ on the basis of assumed
functional forms of $v(r)$, and subsequently we fit the distribution to
the data obtaining a best-fitting $v(r)$.

\subsection{Models of the Systematic Component}

It is non-trivial to write down the mass or velocity profile expected for
the infall region. Most of our current knowledge about the density and
velocity profiles of clusters comes from the inner regions. This is true
both for observational and theoretical work (including N-body
simulations). Simply extrapolating those profiles out to the infall region
might lead to biased results. In the absence of knowledge of the actual
profile, we have found it important to use a number of different model
profiles in order to make sure the results obtained are not just artifacts
of our ignorance.

The adopted profiles should contain as few free parameters as possible,
given the currently meager amount of observational data, the likely
effects of the deviations between our spherical approach and reality, and
the level of ``noise terms" which smear out the clean signal of $v(r)$.

In terms of the ratio of the radial velocity relative to that expected for
a pure Hubble flow at the same distance, $H(r) \equiv v(r)/(H_0r) =
1-v_{peculiar}(r)/(H_0r)$, the first two of our adopted velocity
profiles can be written as:
\begin{eqnarray}
H & = & 1-(\frac{r}{r_{turn}})^{-\beta}\\ H & = & 1 -
\beta \exp(-\frac{r}{r_{turn}}\log\beta ),
\end{eqnarray}
with each of them containing two free parameters which must be estimated.

Two additional profiles were given by specifying the average mass-density
profile as a function of cluster-centric radius:
\begin{eqnarray}
\frac{\overline{\rho}}{\rho_0} & = &
1+(\frac{\overline{\rho}(r_{turn})}{\rho_0}-1)(\frac{r}
{r_{turn}})^{-\gamma}\\ \frac{\overline{\rho}}{\rho_0} & = &
\frac{\overline{\rho}(r_{turn})}{\rho_0}(\frac{r}
{r_{turn}})^{-\gamma}.
\end{eqnarray}
For future reference we label the above four models VI, VII, DI, and
DII, respectively.

As we have seen, for a given value of $\Omega_0$ and $H_0$ there is a
one-to-one correspondence between a velocity profile and a mass profile.
Working directly with a velocity profile emphasizes the fact that
determination of the infall pattern can be considered a purely empirical
task. Furthermore, it can be done independently of $\Omega_0$. However,
for certain combinations of the fitting parameters, the assumed profiles
correspond to unphysical situations. The local density inferred from
the velocity profile may become negative at certain radii within the
region considered. It is straight forward to identify the subsets of
parameters which correspond to unphysical situations, but we shall not
pursue the point any further in this paper. Here we present the method
rather than a determination of real cluster parameters.

The profiles VI, VII, and DI all approach a pure Hubble flow at large
distances, whereas the profile DII does not. Profile DII corresponds to a
situation in which the surroundings of the clusters have a lower
mass-density than the universal average. However, as we consider only the
region near the cluster, one should see the inclusion of profile DII
rather as an attempt to check what will happen if the density is allowed
to vary significantly around the turn-around radius, rather than be forced
to go smoothly towards the background density at the turn-around radius.
While the VI profile corresponds to a faster infall at smaller radii, as
expected for a ``pure" infall situation, the VII profile reaches a minimum
at a certain radius and then approaches zero as the radius decreases. The
latter is in accordance with the behaviour found in N-body simulations,
where (more or less by definition) the radial velocity is zero in the
virial core of the cluster and is followed by a transition zone which can
stretch outside the virial radius (see Figure 5 of Crone et al.\
1994). We expect profile VII to be less sensitive to a misplacement of the
inner cutoff in the data sample, should the cutoff be done at too small an
angular separation. On the other hand, one can obviously not use the part
of the VII profile inside the minimum to determine a density profile for
the cluster, as it becomes difficult to judge how smoothly the mass
estimate produced using the infall profile joins the mass estimate of the
inner, virialized part of the cluster using profile VII.

\subsection{The Distribution Function Corresponding to Purely
Systematic Infall}

Using Figure 1 to define our coordinates, an observer located at a
distance $R$ from the cluster would find the line-of-sight velocity of a
galaxy located at $(\theta,s)$ to be (RG):
\begin{equation} v_{los}
\equiv cz=H_0R\{\cos \theta +(\frac{s}{R}-\cos\theta)H\},
\end{equation}

Let the spatial number-density of galaxies be $n_{gal}(r)$, the number of
galaxies found in a small portion of the ($cz, \theta$)-plane then becomes:
\begin{equation}
n(cz,\theta)dczd\theta = n_{gal}(r)s^2\sin\theta|{\rm
J}|^{-1} dcz d\theta,
\end{equation}
where:
\begin{equation}
{\rm J} =
\left|\matrix{ \frac{\partial cz}{\partial s} & \frac{\partial
\theta}{\partial s}\cr \frac{\partial cz}{\partial \theta} &
\frac{\partial \theta}{\partial \theta} \cr }\right| = \frac{H_0R}{r}
\{ (\frac{s}{R}-\cos\theta)^2\frac{\partial {H}}{H_0 \partial r}+\frac{H}{H_0} 
\frac{r}{R} \}
\end{equation}
is the Jacobian of the transformation between the two coordinate systems.
This result was presented by RG (to within a typographical error). Zero
points in $J^{-1}(cz,\theta)$ make the distribution $n(cz,\theta)$ have a
very characteristic ``horn" shape. Although $n$ naturally has to be
integratable over a finite volume of phase-space, the high-density
regions around these caustic surfaces do contain enough material for the
surfaces to be easily recognisable even in a modest dataset (Figure 2a).
Such a profile was noted by Kaiser (1987) and RG, and its use as a promising
way of determining the infall velocities and enclosed masses as a function
of radius has been investigated (RG, vH).

However, a number of effects complicate the situation making a direct
identification of the ``horn" almost impossible in a set of real
observational data. In most clusters only a small number of galaxies have
been observed within the infall region. A dedicated observational program
can change this picture for the better, but clusters contain a finite
number of galaxies and identifying the ``horn" may demand combining data
for a number of clusters, even if every galaxy in the infall region has a
measured redshift. Secondly, the velocity field is probably not perfectly
radial, even in the region dominated by infall, as envisaged by the
various degrees of substructure on small and intermediate scales seen in
many clusters. That will tend to smear out the caustics (vH), as will
observational errors in the redshifts. In the next section we shall
include a term in the models which incorporates the effects of small-scale
substructure and observational errors.

\subsection{Inclusion of the Noise Term}

Within the hierarchical picture of structure formation the dynamical time
scales for the evolution of small-scale structures (galaxies and groups)
are generally shorter than for larger scales (clusters); a difference
increasing if the power spectrum describing the 
initial density fluctuations on those scales becomes steeper. 
When the dynamical 
time scales are sufficiently 
different, the
internal dynamics of pairs and groups of galaxies in the infall region can
be considered independent of the slower formation of the cluster. If so,
the effect of small-scale substructure on the velocity field of the infall
region of a cluster will be isotropic and independent of the position of
the galaxies with respect to the cluster. We shall assume this to be the
situation, in which case a rough estimate of the effects of small-scale
structure can be obtained by folding the distribution with a Gaussian
smearing function,
$G(x,\sigma)=1/(\sqrt{2\pi}\sigma)\exp(-0.5x^2/\sigma^2)$, leading to:
\begin{eqnarray}
n_{\sigma_{cz}}(cz,\theta)& =&
\int_{-\infty}^{+\infty} n(cz,\theta')G(cz'-cz,\sigma_{cz})dcz' \\ &
=& \int_{-\infty}^{+\infty} n_{gal}(r[\theta,s'])G(cz[\theta,s']-
cz,\sigma_{cz})s'^2\sin \theta ds',
\end{eqnarray}
the latter being the form most easily handled in numerical calculations for
relevant values of $\sigma_{cz}$.

The observational data consist of inhomogeneous datasets obtained
using different selection functions. In general, however, the selection
functions are much nearer to being magnitude limited rather than volume
limited, and we have to include the depth effect, at least for clusters as
near as Coma.

Let $n_L(M)$ be the luminosity distribution of galaxies, and assume $n_L$
is independent of the position (within the infall region) relative to the
cluster position. If the sample is magnitude limited, we can write
the proper distribution function as:
\begin{equation}
n_{\sigma_{cz}}(cz,\theta,m)=\int_{-\infty}^{+\infty}
n_{gal}(r[\theta,s'])
G(cz[\theta,s']-cz,\sigma_{cz})n_L(M[m,s'])s'^2\sin\theta ds'
\end{equation}
and $M(m,s)=m-25-5log_{10}s$, with $m$ and $M$ being apparent and
absolute magnitudes, respectively.

This is our final distribution function. We shall now test its ability to
infer correct parameters on simulated clusters, but first we demonstrate
the effect of $\sigma_{cz}$. Figures 2a--2c show the distribution in 
the $(cz,\theta)$-plane with and without the noise term included. The 
improvement
from adding the noise term is quite evident, when compared to a real case
as in Figure 2b (Coma). Figure 3 shows the number density along a
line-of-sight for various degrees of smearing. Figures 2a, 2c, and Figure 3
were constructed using profile VI, $\beta=1.4$, and $r_{turn}=9Mpc$.

Which value should be used for $\sigma_{cz}$? There are not enough data to
allow $\sigma_{cz}$ to be a free parameter in the models. Rather, we find
guidance from the observed pairwise velocity dispersion (reduced to that
of a single galaxy) at small scales. Unfortunately, there is no current
consensus as to the proper level of the pairwise velocity dispersion. The
recent determination by Markze et al.\ (1995) found $\sigma_{1-2}=540$ km
s$^{-1}$, markedly higher than the previous value, $\sigma_{1-2}=340$ km
s$^{-1}$, found by Davis \& Peebles (1977). However, as shown by Markze et
al.\ (1995), $\sigma_{1-2}$ is strongly influenced by the fraction of
highly clustered galaxies included in a sample, as these galaxies, despite
being relatively few, have a significant effect through their large
relative velocities. Removing all Abell clusters of richness class
$R\ge1$, they find $\sigma_{1-2}=295$ km s$^{-1}$. This could argue for
the use of a low value. However, the current neglect of non-spherical
effects and the future extra noise from superposition of a number of
clusters calls for a value somewhat above the estimate from pairwise
velocity dispersions. Currently we bracket the problem by using the two
values, $\sigma_{cz} =300$ km s$^{-1}$ and $\sigma_{cz}=400$ km s$^{-1}$.

In the above calculations we have adopted $n_{gal}/n_0=1+\xi_{cg}$, with
$\xi_{cg}=(r/r_c)^{-\gamma}$ being the cluster-galaxy correlation function
and using $\gamma = 2.5$ (Lilje \& Efstathiou 1988), $n_0$ is the mean
number-density of galaxies in the Universe. Whereas this is a natural
choice in cases where little is known about the number density profile of
the cluster in question, other choices are of course possible, e.g. using
the observed (deconvolved) profile -- though one should notice that it is
very difficult to determine the luminosity profile of a single cluster in
the region considered here. We have experimented using different
parameterizations of $n_{gal}(r)$. Fortunately, it turns that the method is
rather insensitive to these changes.

\subsection{Fitting a Model to the Data}

We fit the models to the data by maximising the expression:
\begin{equation}
\ln{\rm L}({\rm model}) = \sum_{i=1}^N \ln
p(cz_i,\theta_i,m_i|{\rm model}),
\end{equation}
where:
\begin{equation}
p(cz,\theta,m|{\rm model}) \equiv \frac{
n_{\sigma_{cz}}(cz,\theta,m)}{
\int \int \int n_{\sigma_{cz}}(cz',\theta ',m')dcz' d\theta ' dm'}
\end{equation}
is the probability of observing a galaxy at $(cz,\theta,m)$, given a
particular model and its parameters, and the integral is over the region
of phase space from which the data are drawn. $N$ is the total number of
galaxies in the sample.

This procedure is superior to envelope fitting of the caustics, as all
data points contribute to the fit, thereby making the most of the
available information. Due to the inclusion of the noise term,
$p(cz_i,\theta_i,m_i|{\rm model})$ is a rather gently changing function in
$cz$, and in practice it turns out not to be necessary to discretize
$p(cz,\theta,m)$ in $\theta$ or $cz$ before calculating ${\rm L}$, even
for rather small data samples. One should check, however, that no galaxies
of highly unusual brightness for a sample of a given size and distance are
included when fitting, as such a galaxy can have a large impact on the
results. Such a galaxy should either be excluded or $p(cz,\theta,m)$ has
to be discretized with respect to magnitude.

An important, valuable feature of the model is the fact that it is not
necessary to correct for foreground and background objects, as they are
included in the model. This is not the case for members of neighbouring
clusters however, and care should therefore be taken to select data from a
region of phase-space surrounding each cluster which is small enough that
nearby clusters will only have a minor effect on the velocities. On the
other hand, in the absence of neighbour problems, the method produces
better determinations of the model parameters when the cuts in $cz$ are
sufficiently widespread to include the caustics at all selected $\theta$,
and the cuts in $\theta$ are wide enough to cover the whole region between
the radius of no shell crossing and $r_{turn}$. In each case one has to
weigh these arguments and find a working compromise.

\subsection{Testing the Procedure on Simulated Clusters}

Previously we demonstrated that the inclusion of a noise term in the
spherical infall model and a few further reasonable assumptions produce
simulated clusters which `to the eye' mimic real clusters quite well
(Figure 2).

We shall now test our maximum-likelihood estimator, by using it on
simulated clusters in order to evaluate whether it is capable of
identifying the correct parameters at all (recalling that the previous
attempts to identify caustics for real clusters failed). If so, what
is its sensitivity and how big are the errors arising from assuming wrong
model parameters for a given type of cluster?

We have performed numerous tests on simulated clusters for which all model
parameters were known. In most cases the simulated clusters contained 143
galaxies (sometimes 286) chosen randomly according to the distribution
function $p(cz,\theta,m|model)$ (Equation 18) and having $2<\theta<8$,
4930 km s$^{-1} < cz <$ 8930 km s$^{-1}$, a cluster-centric velocity of
6930 km s$^{-1}$, $\gamma_{cg}=2.5$, and a magnitude cutoff at 15.5. These
parameters may appear odd, but they correspond to those of our prime sample
of real data (Coma).

The results are presented in Table 1a. In the table, column 1 indicates
the systematic component with which the simulated clusters were made,
column 2 gives the velocity dispersion used in the smearing function of
the simulated clusters, columns 3 and 4 give the systematic component and
the velocity dispersion assumed when doing the likelihood analysis, column
5 is the number of simulated clusters on which the analysis was done,
column 6 gives the average value of $r_{turn}$ and its dispersion from the
maximum-likelihood analysis, and column 7 gives the average value of
$\beta$ (or $\gamma$ for the density profile models) and its dispersion.
All simulated clusters made using the profile VI were given parameters:
$(r_{turn}, \beta)=(9,1.4)$, using VII: $(r_{turn}, \beta)=(9,8)$, and
using DI: $(r_{turn},\gamma)=(9.5,2.5)$.

First we note that the adopted maximum-likelihood technique is indeed able
to return the proper parameters, even for $\sigma_{cz}$ as high as 400 km
s$^{-1}$. As might be expected the precision depends on $\sigma_{cz}$, the
match between the model assumed for the simulated cluster (reality), and
the model used for analysis. However, even in cases of mis-match or high
$\sigma_{cz}$ the results are still reasonable.

Figure 4 gives an example of a contour plot for the maximum-likelihood
levels found for a random simulated cluster ($\sigma_{cz}=300$ km
s$^{-1}$). The plot shows $\ln(L)$, the cross identifies the position of
maximum-likelihood, and the levels are separated by ${\rm ln}(2)$ and
constantly decreasing as one gets farther from the peak. There is some
degree of degeneracy between $r_{turn}$ and $\gamma$ (and similarly for
velocity profile models between $r_{turn}$ and $\beta$). This reflects the
fact that for certain correlated changes of ($r_{turn},\gamma$) the
corresponding changes in the velocity profile (and density profile) are
rather small. The uncertainty of an individual ``measurement" in some
cases is rather large, and ideally a number of clusters should be used
when determining model parameters for a set of real data.

\subsection{Extension of the Model to Include Distance Information}

Obviously one cannot determine the parameters of an infall profile, $v(r)$
without knowledge of both $v$ and $r$. In the model above that information
comes about rather indirectly in the form of distribution functions
calculated on the basis on an assumed profile $v(r)$ (and other model
parameters) and the observed values of $cz$ and $\theta$. Due to the
existence of multiple solutions for the distance for certain combinations
of $(cz,\theta)$ even without noise, and due to the assumption of a random
scatter in $cz$ of the order $\sigma_{cz}\ge 300$ km s$^{-1}$, the probability
distribution, $p(s|cz,\theta,m)$, of galaxy being at a particular distance
given the observed $(cz,\theta,m)$ is very broad. For most galaxies it is
even impossible to tell whether they are in front of or behind the
cluster. But as has been demonstrated it is still possible to reliably
estimate the parameters of the infall profile.

It seems worthwhile to test whether use of observed distances leads to an
improved determination of the model parameters, even though the
uncertainty in distance estimate for galaxies is at present quite
considerable relative to the sizes of the clusters in question. (If that
was not the case the whole procedure presented in the previous sections
clearly becomes obsolete, and one could do much better using another
approach.)

Therefore, assume that for a given distance estimator the probability that
a galaxy of true magnitude $M_t$ and inferred magnitude $M_{o}$ is given
by $G(M_o-M_t,\sigma_m)$. For the Tully-Fisher relation $\sigma_m$ is of
the order 0.38 (c.f. Willick et al.\ 1997). With $s$ being the true
distance, the probability of observing a galaxy at distance $s_o$ becomes:
\begin{equation} p(s_o|s)=\frac{5}{\ln10\sqrt{2\pi}\sigma_m}\frac{1}{s_o}
e^{-\frac{25log_{10}^2(s_o/s)}{2\sigma_m^2}}.
\end{equation}

Including this term in the expression for $n(cz,\theta,m)$ (Equation 16
and further) provides a simple way of including the distance information
in the likelihood estimates. (As the above expression is already
normalized and does not depend on model parameters other than $\sigma_m$,
the extra integration over $s_o$ in the normalization of
$n(cz,\theta,m,s_o)$ is trivial as long as no cuts are made in $s_o$ space
when selecting the sample.)

Table 1b shows the result of including distance information in the tests
on simulated clusters. Except for the added distance information (assuming
$\sigma_m=0.38$) for the simulated clusters and the inclusion of the
distance term in the distribution function used in the maximum-likelihood
analysis, everything is performed exactly as outlined in the previous
sections, including the format of the table. Clearly distance information
has a significant impact even for realistic values of $\sigma_m$. It
would be of great value to have distances for the galaxies which are to be
used in deriving infall profiles, despite the $\sim$20\% uncertainty of a
single distance measurement. It is not clear whether the best observing
strategy is to start measuring more distances to galaxies in the infall
regions of a few clusters or to measure many more redshifts of galaxies in
a larger number of clusters.

\subsection{Combining Data from Different Clusters or Datasets}

As long as the parameters to be determined can be assumed the same within
different clusters or datasets the information can be combined by simply
adding the likelihoods found for the individual datasets before searching
the best fitting parameters. It is thus possible to combine, for example,
a dataset with distance information with a larger one without distance
information for a given cluster, even if they were made using different
selection criteria or to combine different clusters observed using
different selection criteria. Clearly the galaxies within a given dataset
must have been chosen randomly from within the redshift, angular
separation, and magnitude bins describing that dataset.

If the clusters in question are not ``equal", which in general will be the
case, it becomes necessary to rescale them before performing the
maximum-likelihood analysis. This can be done easily and with good
results, as demonstrated by Carlberg et al.\ (1997b), who studied the
virialized regions.

\section{RESULTS FOR COMA}

We shall now apply the models presented above to the Coma cluster. As
mentioned above one should be cautious when using data for one cluster
only. Unfortunately, Coma is currently the only cluster for which we have
access to the type of data needed for a large part of the infall region.

Since the model includes contributions from background galaxies, but not
from other clusters, it is important to select a region of phase space
which is not `contaminated' by other clusters, yet contains as many
galaxies belonging to the infall region of Coma and its outskirts as
possible and simultaneously is large enough to constrain the sought-after
parameters efficiently.

Figure 5 shows Coma viewed in the $(cz, \theta)$-plane. Filled circles are data
from {\it zcat} (Huchra et al 1993) and open triangles are from vH. The rectangle
inside the dashed lines defines the region of phase space we decided to
model. It is a bit narrower in $cz$ than we would like, but this was done
in order to avoid including a significant number of galaxies from the
`filament' extending from the cluster A1367. 

The inner cutoff has to be large enough so that virialization and shell
crossing have no effect on the galaxies selected. We use
$\theta_{low}=2^{\circ}$ which corresponds to $r\approx2.4h^{-1}Mpc$. The
outer cutoff is 8$^{\circ}$, large enough to contain most of Coma, yet
small enough to avoid inclusion of galaxies heavily interacting with the
surrounding clusters. The precise values of the two cutoffs are not
important, and we have checked this by also running our models with
cutoffs at 3$^{\circ}$ and 10$^{\circ}$.

It is well known that Coma is not perfectly spherical and that its early-
and late-type galaxies are not distributed similarly. This has been
demonstrated convincingly, for example, by Doi et al., (1995). To
study the effects of this substructure in Coma on our analysis, we
subdivided the dataset in three different ways: firstly into early- and
late-type galaxies (according to Hubble type), secondly into inner and
outer galaxies (according to $\theta$), and thirdly in high- and
low-density regions, according the whether or not the galaxies are near or
far from the two filaments (in $\phi$) visible in Figure 6. In the latter
two cases the cuts were made such that each splitting resulted in
approximately the same number of galaxies in the two parts.

Despite going to much fainter luminosities, there is a distinct drop in
the number of galaxies per magnitude bin at 15.5 in the full sample.
Therefore, we have made a lower cutoff in luminosity at m=15.5.
Furthermore, we decided to ignore the extra data included in the vH
sample, as only few of them are found within the region of phase space
considered here.

The results are presented in Tables 2 and 3, while Figures 7a -- 7c are
examples of the likelihood contours found (with symbols as in Figure 4).
In Table 2 the left half shows results for the DI model, while the right
half are results for the DII model. The fits are for all 143 galaxies in
the dataset. The value of $\Omega_0$ shown on the left is the one assumed
in the analysis in order to convert a mass profile to a velocity profile.
In Table 3 the left half contains results for the VI velocity profile, the
right half for the VII velocity profile. The uppermost line gives the
result for all 143 galaxies, whereas the results below are for the various
subdivisions described above.

We note how robust the estimates of the turn-around radius are to changes
in the assumptions concerning $\Omega_0$, $\sigma_{cz}$, and the velocity
profile. For the total dataset the full range of values returned is
$9.0h^{-1}Mpc \le r_{turn} \le 10.0 h^{-1}Mpc$. Table 2 illustrates the
effect of changing $\sigma_{cz}$. An increase in $\sigma_{cz}$ will
increase the width of the caustics for a given velocity profile, hence a
fixed dataset results in a lower estimate of the systematic infall 
velocity at a given distance from the cluster.

The only subsets of Coma data which result in a clear change in the
velocity profile are the ``high-density" galaxies and the ``inner"
galaxies. These are exactly the subsets that we would expect to deviate
{\it a priori} from the analysis, the high density galaxies because local
effects are likely stronger in the dense surroundings and because the
marked deviation from our assumption of spherical symmetry (e.g.
$n_{gal}(r)$ being spherically symmetric). The inner galaxies deviate
because the strongest signal in that region comes from the galaxies
falling fastest towards the cluster, thus providing little information
towards a determination of the turn-around radius.

\subsection{Implications of the Results for Coma}

What can we deduce from the Coma results? Firstly, we do not have enough
data to arrive at any firm conclusions. The parameters found are not
strongly constrained, and working on a single cluster our assumption of
spheridicity might well break down.

Having said that it should be noted that the velocity profiles found are
very similar for a given value of $\sigma_{cz}$. Not only is this true
among models and cosmologies, interestingly enough it is also the case for
the subdivided sets of Coma, with the exception of precisely those sets
for which we would expect deviations. It seems our results are not
completely arbitrary!

As a further check we can compare the mass estimates provided by our
models at a small radius to the virial estimate of the mass of the Coma
cluster. This is done in Figure 8, where we plot the estimated masses at
$r=2.5h^{-1}Mpc$ and $r_{turn}$ using the DI profile results. Circles are
for $\sigma_{cz}=300$ km s$^{-1}$, triangles for $\sigma_{cz}=400$ km
s$^{-1}$, and filled symbols for the smaller of the two radii. The inner
mass estimates are nearly insensitive to the adopted value of $\Omega_0$,
as one would expect. The mass estimates at $2.5h^{-1}Mpc$ are around
$2\times10^{15}h^{-1}M_\odot$ for the low-velocity dispersion models and
around $1.5\times10^{15}h^{-1}M_\odot$ for the high-velocity dispersion
models. Based on X-ray data Hughes (1989) found
$M(r\le2.5h^{-1}Mpc)=0.5-1.5\times10^{15}h^{-1}M_\odot$, with preferred
values in the range
$M(r\le2.5h^{-1}Mpc)=0.8-1.1\times10^{15}h^{-1}M_\odot$, while Watt et
al.\ (1992) estimate the total mass to be $1.3\times10^{15}h^{-1}M_\odot$.
Using galaxy line-of-sight velocities, The \& White (1986) find $M(r\le2.7
h^{-1}Mpc)=0.3-2.5\times10^{15}h^{-1}M_\odot$, with a value of
$0.95\times10^{15}h^{-1}M_\odot$ assuming mass and galaxies are similarly
distributed. Colless \& Dunn (1996) find
$M_{virial}=0.9\times10^{15}h^{-1}M_\odot$. As emphasized by these and
other authors, determinations based solely on galaxy velocities yield
large uncertainties. Recall that our results do not rely in any way on
data from the virialized region (except for providing the center of Coma).
The good agreement between our results and the results quoted above is
strong evidence that the method works, notwithstanding our ``testing"
laboratory with its simulated clusters. Performing a similar analysis, but
using results found for slightly different cuts in angular selection
(3--8$^{\circ}$, 2--10$^{\circ}$, 3--10$^{\circ}$), we find
$M(r\le2.5h^{-1}Mpc) = 1.1-2.4\times10^{15}h^{-1}M_\odot$, again
indicating that our results are far from accidental.

If we had good knowledge of the luminosity profile of Coma, its M/L ratio
as a function of radius could be derived for the whole infall region when
combined with our results. Unfortunately, such a profile does not yet
exist. vH did a very careful study of the light profile of Coma, using Ca
and APM data separately to derive two light profiles, but they are based
solely on data from the region $\theta \le 2^{\circ}$, making
extrapolation to $r_{turn}\approx 10 h^{-1}Mpc$ too uncertain.

Alternatively, we attempt to use the luminosity profile, $n_{gal}(r)$,
specified in the models. Kent \& Gunn (1982) estimated the total
luminosity of Coma inside $\theta \le 3^{\circ}$ to be
$L_B=4\times10^{12}h^{-2}L_\odot$. This value can be used for
normalization of our luminosity profile, and we can derive estimates of
the total luminosity inside a given radius. As an example, Figure 9 shows
the mass-to-light ratio calculated this way at $r=2.5h^{-1}Mpc$ and at
$r_{turn}$ (filled versus open symbols) for the DI profile for both the
low and high value of the velocity dispersion parameter (circles and
triangles).

In the slightly non-linear regime one has:
\begin{equation}
\frac{v(r)}{H_0r}=1-\frac{1}{3}\Omega_0^{0.6}\frac{\Delta}{(1+
\Delta)^{0.25}},
\end{equation}
where $\Delta$ is the relative mass over-density (Yahil 1985). That the
luminosity does not necessarily follow the mass can be taken care of
through a bias parameter: $\Delta_L= b\Delta$. In this case the peculiar
velocity scales approximately as $\Omega_0^{0.6}/b^{0.75}$ when $\Delta_L$
is substituted directly for $\Delta$ in Equation 20, in the outer infall
region where the average over-density is larger than one, yet small enough
that Equation 20 still provides a good fit to the real non-linear infall
solution (see RG Figure 5).

In Figure 10 we show some of the velocity profiles obtained using our
model as well a set of curves based upon Equation 20 and using
$n_{gal}(r)$ to provide an estimate for $\Delta_L$.  The thick solid lines
correspond to predicted velocity profiles based on the light profile
assuming $\Omega_0^{0.6}/b^{0.75}$ = 1.0, 0.8, 0.6, 0.4, and 0.2, where
1.0 is the lowest line. The thin solid lines are from models VI and VII,
the thin dashed lines the DI profile for both values of $\sigma_{cz}$, and
the thin long-dashed lines are for the DI profile as above, but for
$\theta$ in the 3$^{\circ}$ -- 10$^{\circ}$ interval. From Figure 10 we
see that  $0.6 < \frac{\Omega_0^{0.6}}{b^{0.75}} < 0.8$, but we caution
the reader that the basis for the luminosity profile we use, the
cluster-galaxy correlation function, is not necessarily a very precise
description of the light profile of Coma throughout the infall region.
While it serves in our models as a reasonable luminosity profile (since
the models are not very sensitive to this), one should be careful in
interpreting variations of the M/L ratio of Coma based on this profile.
Also recall that our mass estimates for the inner region were somewhat too
high, indicating the estimated infall velocities could be systematically a
bit too large. 

Currently much work is being done on the relation between cluster profiles
and cosmology (e.g. Cole \& Lacey 1996, Navarro, Frenk, \& White 1997).
Nearly all the work, however, applies mainly to the virialized part of the
clusters. It would be very useful to extend the work to the infall region,
in which case the density profiles derived using models such as those
presented here could be compared with results from N-body simulations.

The values of $r_{turn}$ for Coma inferred from our models show little
variance. Given some well-defined inner radius (e.g. the virial radius,
$r_{vir}$) or the radius at a given average over-density, one might
intuitively expect that the ratio $r_{turn}/r_{vir}$ would allow
discrimination between different cosmological models. There is an
indication from Figure 5 of Crone et al.\ (1994) that this may be the case,
but it is probably only through future high-resolution N-body simulations
that one can hope to answer such a question.

\section{DISCUSSION}

The main difference between the models presented here and previous
attempts to determine the masses of clusters on the basis of their infall
velocity profiles is the inclusion of a noise term, which represents the
effects of small-scale structure and observational errors, and the use of
the full distribution function, $f(cz,\theta,m)$, rather than the caustic
envelope, when fitting models to data. Furthermore, our modelling does not
rely strongly on the light profile, which is not well determined in the
infall region of individual clusters.

Visual inspection reveals that in the limit of a no-noise term the model
distribution and the real distributions are dissimilar (Figure 2). With
hindsight it is therefore not surprising that RG and vH were unable to
decide upon a best-fit model. Had they found one it might have resulted in
false conclusions, as the effect of small-scale substructure and
observational errors, represented through $\sigma_{cz}$ in our models, is
to increase the apparent width of the caustic surfaces. Without accounting
for this effect the inferred systematic velocities become too high.

With the noise term included in our models, the $(cz, \theta, m)$
distributions become acceptable to the eye (Figure 2), but as discussed
earlier such a smearing of the caustics is also expected from an
aspherical infall. Which of the two interpretations is correct? Most
likely a combination, as both explanations are based upon known physical
effects. It is well known that clusters are not perfectly spherical (see
Figure 6 of the Coma cluster) which must result in a systematic flow which
is not perfectly radial if the distribution of the dark matter is anywhere
near the distribution of the galaxies. On the other hand, galaxies do
induce substantial non-radial motions upon one another, even on scales so
small that the effect clearly is not accounted for by a non-spherical
cluster potential (e.g. in pairs and groups of galaxies) which should be
combined with an observational error of the order $\sim$100 km s$^{-1}$.
Furthermore, the level of $\sigma_{cz}$ which leads to models resembling
Coma is of the order 300--400 km s$^{-1}$, which is very close to the
observed pairwise velocity dispersion for a single galaxy, which we expect
to be a good measure of the ``noise" on small scales.

Are the effects of overall non-sphericity, be it a non-spherical potential
resulting in a non-radial flow or a radial flow but mainly along a few
filaments, so strong that they hamper the use of the spherical infall
model, or the use of the particular models presented here? There is no
doubt that using the spherical approximation for individual clusters one
risks drawing false conclusions, as demonstrated by
van Haarlem \& van de Weygaert 1993 on N-body data. They found everything
from good to poor agreement between a spherical infall solution and
the actual radial velocity distribution of the 
various clusters in their simulations. Thus to model individual clusters one
will in general need to go beyond the spherical approximation. An attempt to
do that has been made by Diaferio \& Geller (1997).

But evidently
an ensemble cluster, made by the superposition of a number of clusters,
must be spherical.  This is found to be the case for
clusters in N-body models (van Kampen, 1995). As one does not have spatial
3D information about real clusters, one might worry that selection
effects makes an "observed" ensemble cluster non-spherical, but rather
elongated along the line of sight.  We believe that such problems can 
be minimized by using an X-ray selected cluster sample. Work is in progress,
using the simulations by van Kampen (1995), to quantify to which degree there
might be a problem if using an ensemble of clusters selected on the
basis of "optical" data.

The ensemble cluster being spherical, the relevant question 
is whether the systematic
flow of the galaxies in the ensemble cluster can be described by the
spherical approximation and related to cosmology. Lilje \& Lahav (1991)
found this to be the case. They compared the predictions of the spherical
infall model applied to density peaks in a random Gaussian field to the
outcome of N-body simulations and found the analytic models to give a good
representation of the density and velocity profiles found in the
simulations.

We conclude that models like the ones presented here will work correctly
when applied to a few clusters simultaneously. Further evidence comes from
the mass estimates of the virial region of Coma produced with our infall
method which are fairly close to the estimates produced based on
observations of the inner region of Coma.

How superposition of cluster data will effect the best value of
$\sigma_{cz}$ to use in the models is hard to say. On the one hand, one
might expect that if non-radial flows are significant in individual
clusters it will lead to an increase in the best value to use when
considering a superposition. On the other hand, the values of
$\sigma_{cz}$ might be high already. Assume, for example, that deviations
from spherical symmetry smear out the infall profile of Coma. Then the
fact that our models represents Coma quite well could indicate that the
level of $\sigma_{cz}$ adopted is higher than what corresponds solely to
small-scale substructure. Hopefully, future observational work on the
level of the pairwise velocity dispersion on small scales can help answer
such questions. It is an important parameter in the models, and it should
ideally be determined without using the models themselves. For the purpose
of these models one needs estimates of $\sigma_{pairwise}$ which excludes
the central regions of clusters, as we need the velocity dispersions on
small scales in the probably much ``colder" infall regions.

The best way to improve the models of the infall region is, we believe, 
by means of studying N-body simulations with emphasis on this region. As the
typical systematic velocities in the infall region can easily be as small at
the internal velocity dispersions of galactic dark matter halos particular care
has to be taken the problem of identifying "galaxies" and measuring the
velocity field of only those becomes critical.  Simulations including
simplified galaxy formation, like the ones done by van Kampen (1995), seem
an ideal basis for such a study.
Potentially N-body models can provide knowledge about eventual systematic 
effects
such as asphericity and about the way in which $\sigma_{cz}$ might change
due to superposition of clusters, or with the position relative to the
cluster center. Also, given that we have found that the value of the
turn-around radius of Coma is very insensitive to our modelling, it would
be of great interest to identify in N-body simulations any systematic
relation between the outer properties of clusters (such as the turn-around
radius) and the inner properties (parameters which can be determined on
the basis of virial estimates or similar) which relate to cosmology.

\section{CONCLUSIONS}

We have modelled the velocity field of the infall region of clusters of
galaxies by a superposition of a systematic component, which is
spherically symmetric and based on the spherical infall approximation, and
a noise component, which is isotropic and independent of position in the
cluster, which represent the combined effects of small-scale substructure
and observational errors. These random velocities are assumed Gaussian and
their level described by $\sigma_{cz}$. In the models the mass-to-light
ratio is allowed to vary freely. Knowing the systematic component one can
deduce the velocity and mass profile of the clusters.

For $\sigma_{cz}=300-400$ km s$^{-1}$ our models resemble Coma quite well.
This level corresponds to the observed pair-wise velocity dispersion of
galaxies on small scales. The effect of the noise term is to smear out the
caustic surfaces searched for by previous groups. The models allow
calculation of the phase-space density of galaxies expected at any point
in the observable $(cz, \theta, m)$-space outside the region of shell
crossing. The likelihood of obtaining a given set of data from a
particular model and its parameters can be determined. It is demonstrated
on simulated data the maximum-likelihood procedure allows one to identify
the correct model parameters, even for realistic noise levels and sample
sizes.

To avoid the effects of deviations from sphericity and to enhance the
likelihood signal, data for a few clusters should be superimposed when
applying the model to real observations. This is technically a simple
process, but currently sufficient observational data are not available.

The models do not in any way rely on data for the virial region of the
clusters, other than providing the center of the clusters. Hence, the
model determination of the cluster mass at small radii can be compared to
the virial estimates as a test of the method.

Finally, it is demonstrated that the inclusion of distance information of
the precision which can be obtained using the Tully-Fisher relation
strengthens the likelihood signal significantly. Potentially this could be
used to allow deviation from the assumption of spherical symmetry.

We have applied our models to the only cluster for which we have enough
data far out in the infall region, namely Coma. The estimated mass at
$r=2.5h^{-1}Mpc$ falls in the range $1.1-2.4\times10^{15}h^{-1}M_\odot$,
depending on the model and the level of the noise term. Higher
$\sigma_{cz}$ produces a lower mass estimate. We find the turn-around
radius, $r_{turn}$, to be of the order $10 h^{-1}Mpc$, which is fairly
insensitive to changes in the models, the noise level, and even to
subdivision of the data. 

Using the cluster-galaxy correlation function to provide an estimate of
the light profile, we find for Coma  $0.6 <
\frac{\Omega_0^{0.6}}{b^{0.75}} < 0.8$, $b$ being the bias parameter:
$b\Delta_{mass}=\Delta_{light}$, where $\Delta$ is the relative
over-density. We hesitate to draw any firm conclusion on the basis of this
result, as long as the present analysis was performed on data from only a
single cluster. 

However, we have demonstrated that the method does indeed work, and it
provides a promising way of measuring the infall velocities in the outer
regions of clusters. More observations of redshifts in the outskirts of
clusters are needed as well as a dedicated study of the properties of the
infall region of clusters from N-body simulations.

\acknowledgements

It is a pleasure to thank C. Pritchet and J. Poll for stimulating
discussions during the early phases of this work, much of which was
carried out when H.V.\ was a postdoc at University of Victoria. H.V.\
acknowledges support from the Danish Research Council for Natural Sciences
which made this possible, and from Danmarks Grundforskningfond
through its support for an establishment of the Theoretical Astrophysics
Center. F.D.A.H.\ gratefully acknowledges financial support through an
operating grant from NSERC. We thank A. P. Cowley for a careful
reading of the manuscript, and finally we wish to thank our anonymous referee 
for comments which let to an improved paper.

\clearpage

\begin{table}
\tablecaption{hhhh}
\begin{center}
\begin{tabular}{lccccccc}
\multicolumn{8}{l}{{\bf Table 1a.}
Fits to Simulated Clusters. } \\
\tableline
\tableline
Simulated & $\sigma_{cz}$ & Fit & $\sigma_{cz}$ & \#clu & \#gal/clu  &
$r_{turn}$ & $\beta$ or $\gamma$\\
& [km s$^{-1}$] & & [km s$^{-1}$] & & & [Mpc] & \\
\tableline
DI & 300 & DI & 300 & 300 & 143 & 9.70$\pm$1.69 & 2.53$\pm$0.52 \\
DI & 300 & DI & 300 & 150 & 286 & 9.64$\pm$1.17 & 2.50$\pm$0.37 \\
DI & 400 & DI & 400 & 300 & 143 & 9.59$\pm$2.05 & 2.41$\pm$0.61 \\
DI & 400 & DI & 400 & 150 & 286 & 9.69$\pm$1.41 & 2.45$\pm$0.41 \\
DI & 400 & DI & 300 & 300 & 143 & 9.38$\pm$2.73 & 2.67$\pm$0.73 \\
DI & 300 & DI & 400 & 300 & 143 & 9.93$\pm$1.50 & 2.27$\pm$0.47 \\
VI & 300 & VI & 300 &  70 & 143 & 9.01$\pm$1.65 & 1.44$\pm$0.33 \\
VI & 400 & VI & 400 & 100 & 143 & 8.93$\pm$2.29 & 1.39$\pm$0.35 \\
VII &300 & VII& 300 &  20 & 143 & 9.20$\pm$1.23 & 8.31$\pm$2.51 \\
VI & 300 & VII& 300 &  20 & 143 & 9.49$\pm$1.80 & 7.51$\pm$3.00 \\
VI & 400 & VI & 300 & 100 & 143 & 8.47$\pm$2.73 & 1.57$\pm$0.38 \\
VI & 300 & VI & 400 & 100 & 143 & 9.70$\pm$1.73 & 1.10$\pm$0.37 \\
\tableline
\end{tabular}
\end{center}
\end{table}

\newpage
\begin{table}
\begin{center}
\begin{tabular}{lccccccc}
\multicolumn{8}{l}{{\bf Table 1b.} Fits to Simulated Clusters, Including
Distance Information.}\\
\tableline \tableline
Simulated & $\sigma_{cz}$ & Fit & $\sigma_{cz}$ & \#clu & \#gal/clu &
$r_{turn}$ & $\beta$ or $\gamma$\\
& [km s$^{-1}$] & & [km s$^{-1}$] & & & [Mpc] & \\
\tableline
DI & 300 & DI & 300 & 300 & 143 & 9.51$\pm$0.91 & 2.51$\pm$0.33 \\
DI & 300 & DI & 300 & 150 & 286 & 9.52$\pm$0.60 & 2.51$\pm$0.23 \\
DI & 400 & DI & 400 & 300 & 143 & 9.33$\pm$1.04 & 2.49$\pm$0.38 \\
DI & 400 & DI & 400 & 150 & 286 & 9.41$\pm$0.69 & 2.49$\pm$0.25 \\
DI & 400 & DI & 300 & 300 & 143 & 9.43$\pm$1.26 & 2.68$\pm$0.41 \\
DI & 300 & DI & 400 & 300 & 143 & 9.37$\pm$0.83 & 2.35$\pm$0.31 \\
VI & 300 & VI & 300 & 70  & 143 & 8.99$\pm$0.81 & 1.42$\pm$0.17 \\
VI & 400 & VI & 400 & 100 & 143 & 8.78$\pm$1.17 & 1.39$\pm$0.23 \\
VII &300 & VII& 300 &  20 & 143 &8.89$\pm$0.99 & 8.90$\pm$2.38 \\
VI & 300 & VII& 300 & 20  & 143&10.40$\pm$0.87 & 5.28$\pm$1.31 \\
VI & 400 & VI & 300 & 100 & 143 &8.95$\pm$1.44 & 1.53$\pm$0.25 \\
VI & 300 & VI & 400 & 100 & 143 &8.56$\pm$0.86 & 1.25$\pm$0.21 \\
\tableline
\end{tabular}
\end{center}
\end{table}

\begin{table}
\begin{center}
\begin{tabular}{lcccccccc}
\multicolumn{9}{l}{{\bf Table 2.} Fits of D1 and D2 to Coma.}\\
\tableline \tableline
 & $\sigma_{cz}$= & 300 km s$^{-1}$  & $\sigma_{cz}$= & 400 km s$^{-1}$
 & $\sigma_{cz}$= & 300 km s$^{-1}$  & $\sigma_{cz}$= & 400 km s$^{-1}$ \\
$\Omega_0$ & $\gamma$ & $r_{turn}$ & $\gamma$ &$r_{turn}$
& $\gamma$ & $r_{turn}$ & $\gamma$ &$r_{turn}$\\
 & & $[Mpc~h^{-1}]$  & & $[Mpc~h^{-1}]$
 & & $[Mpc~h^{-1}]$  & & $[Mpc~h^{-1}]$ \\
\tableline
0.01& 2.8 & 9.5 & 2.5 & 10.0 & 2.8 &  9.5   & 2.5 & 10\\
0.2 & 2.6 & 9.5 & 2.4 & 10.0 & 2.8 &  9.0   & 2.5 & 9.5\\
0.4 & 2.5 & 9.5 & 2.3 & 10.0 & 2.6 &  9.0   & 2.4 & 9.5\\
0.7 & 2.5 & 9.5 & 2.2 & 10.0 & 2.7 &  8.5   & 2.4 & 9.0\\
1.0 & 2.4 & 9.5 & 2.2 & 10.0 & 2.6 &  8.5   & 2.3 & 9.0\\
\tableline
\end{tabular}
\end{center}
\end{table}

\begin{table}
\begin{center}
\begin{tabular}{lcccccc}
\multicolumn{7}{l}{{\bf Table 3.} Fits of V1 and V2 to Coma.}\\
\tableline \tableline
Component & $\theta$ & N & $r_{turn}$ & $\beta$ & $r_{turn}$ & $\beta$ \\
                & degrees &     &  $[Mpc~h^{-1}]$  &        &
$[Mpc~h^{-1}]$ & $[Mpc~h^{-1}]$ \\
\tableline
All             &  2-8    & 143 &  9.5  & 1.6  & 10.0  & 9.0  \\
Early types     &  2-8    &  51 &  9.0  & 1.7  & 10.0  & 8.0   \\
Late types      &  2-8    &  92 & 10.0  & 1.7  & 10.5  &10.5   \\
High density    &  2-8    &  74 & 13.5  & 1.4  & 14.0  & 7.0  \\
Low density     &  2-8    &  69 &  9.5  & 1.4  & 10.0  & 6.5  \\
Inner region    &  2-5    &  73 &  8.0  & 2.0  &  8.5  &12.0   \\
Outer region    &  5-8    &  70 & 10.5  & 2.0  & 10.5  &12.0  \\
\tableline
\end{tabular}
\end{center}
\end{table}

\clearpage


\begin{figure} 
\begin{center}
\epsfig{figure=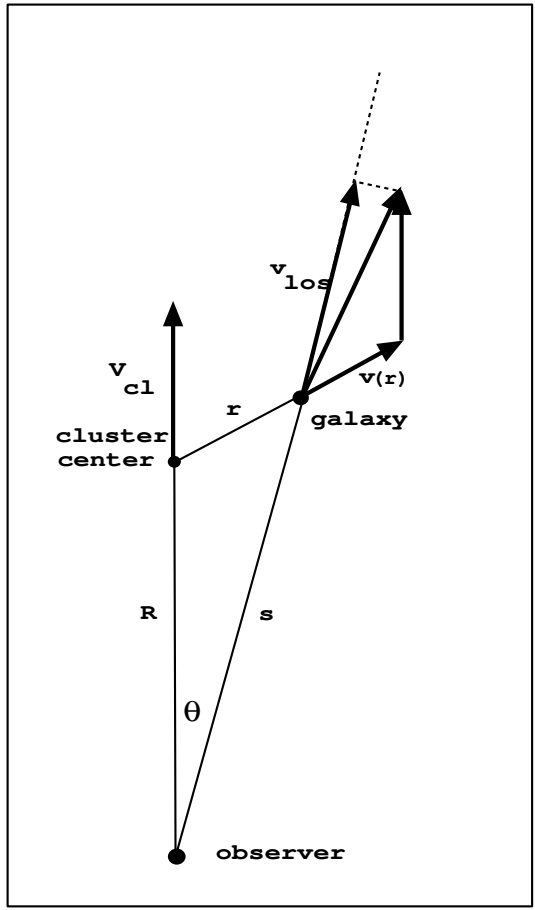,height=12.0cm}
\end{center}
\caption{A diagram showing how the various geometrical quantities in the
models are defined.}
\end{figure}

\begin{figure}  
\begin{center}
\epsfig{figure=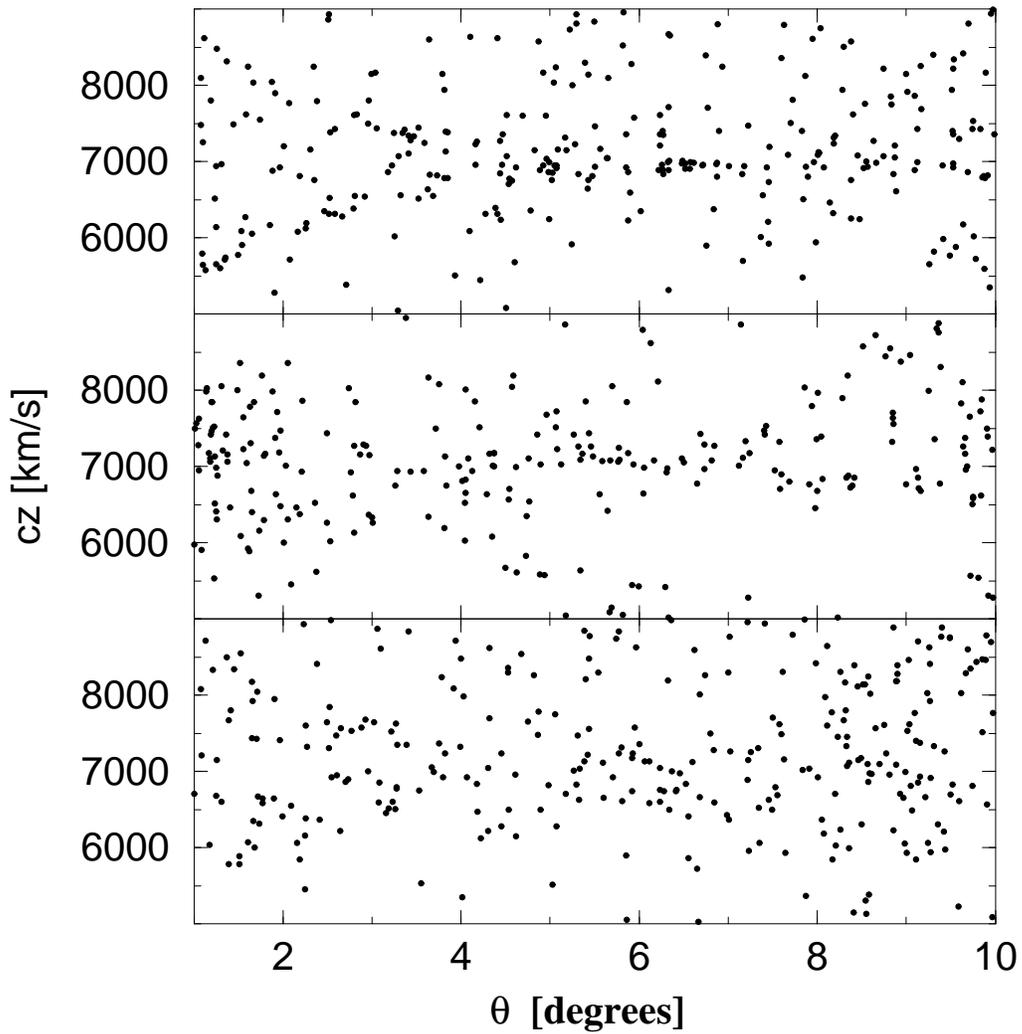,angle=270}
\end{center}
\caption{($\theta,cz$) diagrams of, (a) model based on profile VI with
$r_{turn}=9, \beta=1.4$ and no smearing term (Equation 12 but including
magnitude term as in Equation 16), (b) Coma in similar diagram (only
galaxies having apparent magnitude $<15.5$), and (c) as (a) but now
including a smearing term in $cz$ of strength $\sigma_{cz}=300$ km s$^{-1}$
(Equation 16).}
\end{figure}

\begin{figure}  
\begin{center}
\epsfig{figure=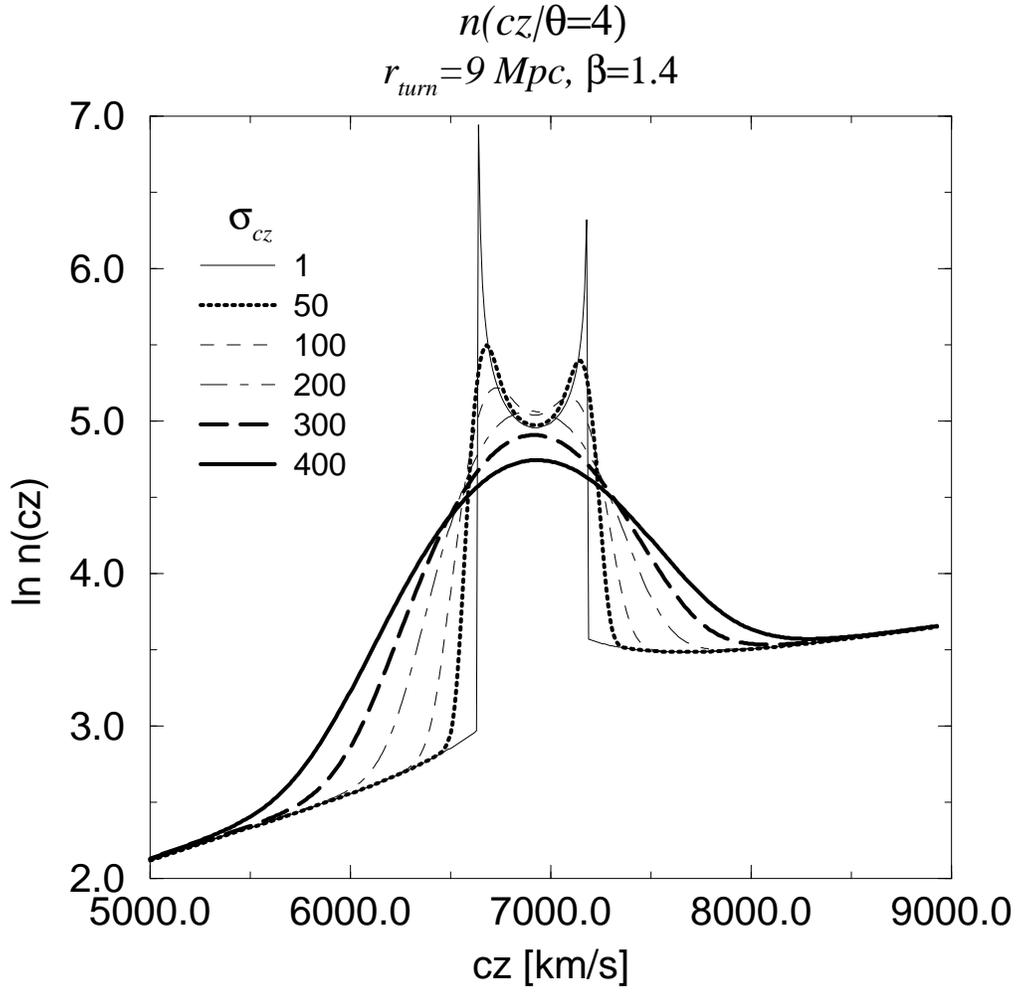,angle=270}
\end{center}
\caption{$n(cz|\theta,\sigma_{cz})$ curves for $\theta= 4^{\circ}$ and
different values of $\sigma_{cz}$.}
\end{figure}

\begin{figure}  
\begin{center}
\epsfig{figure=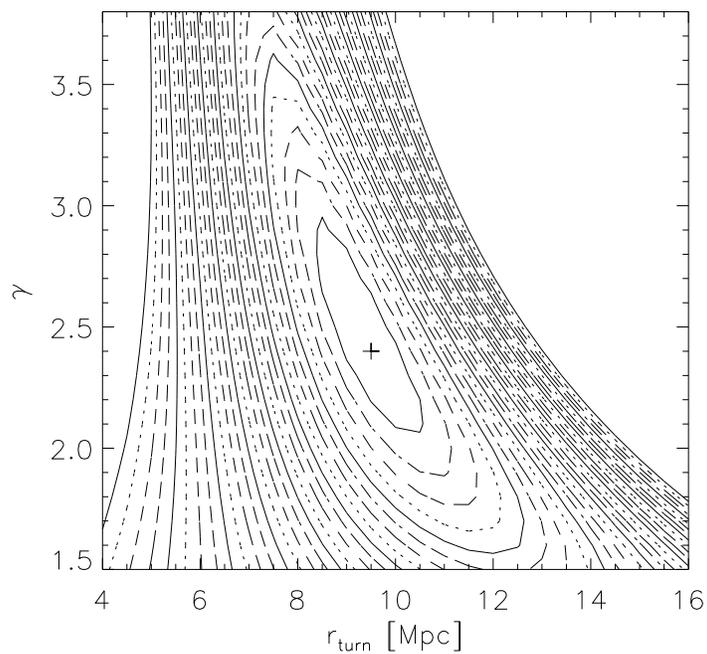}
\end{center}
\caption{Likelihood contours for a simulated cluster constructed using
model DI with $\gamma=2.5$, $r_{turn}=9.5$, and $\Omega_0=0.4$. The
cluster contains 143 galaxies and was analysed assuming model DI with
$\Omega_0=0.4$. Shown is ln(L), and contours are separated by ln(2),
maximum is indicated by cross and contours are constantly decreasing
moving away from maximum.}
\end{figure}

\begin{figure}  
\begin{center}
\epsfig{figure=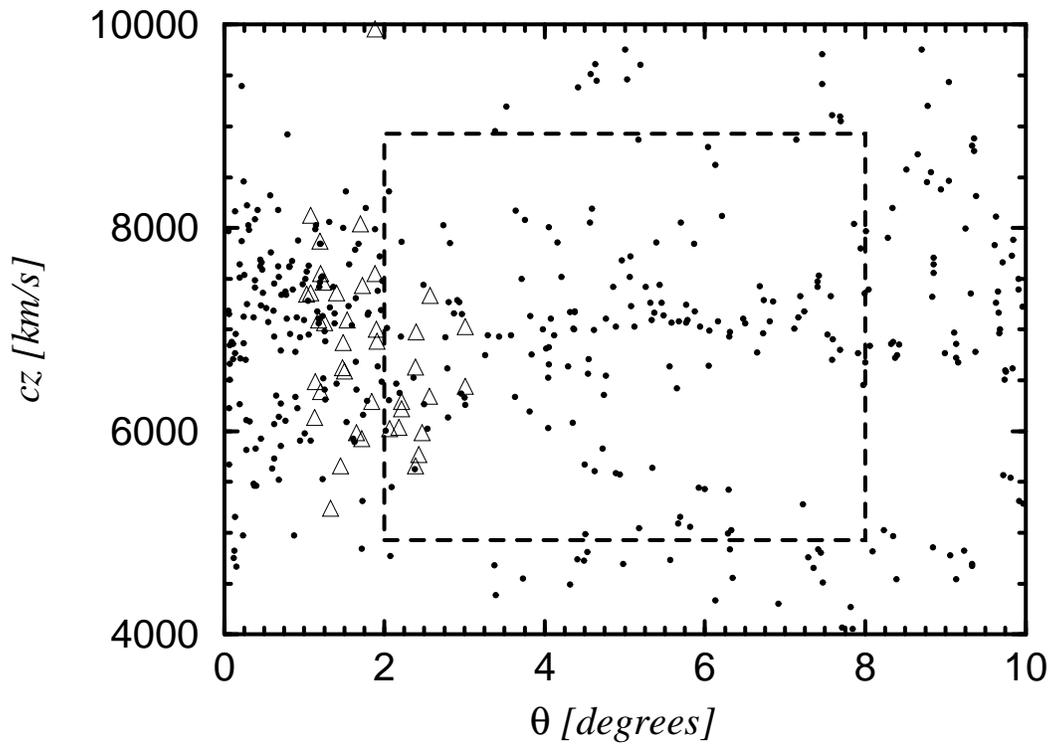,angle=270}
\end{center}
\caption{($\theta,cz$) diagrams of Coma, including data from both {\it
zcat} and vH. Region surrounded by dashed lines is the one selected for
maximum-likelihood analysis. See text for details.} 
\end{figure}

\begin{figure}  
\begin{center}
\epsfig{figure=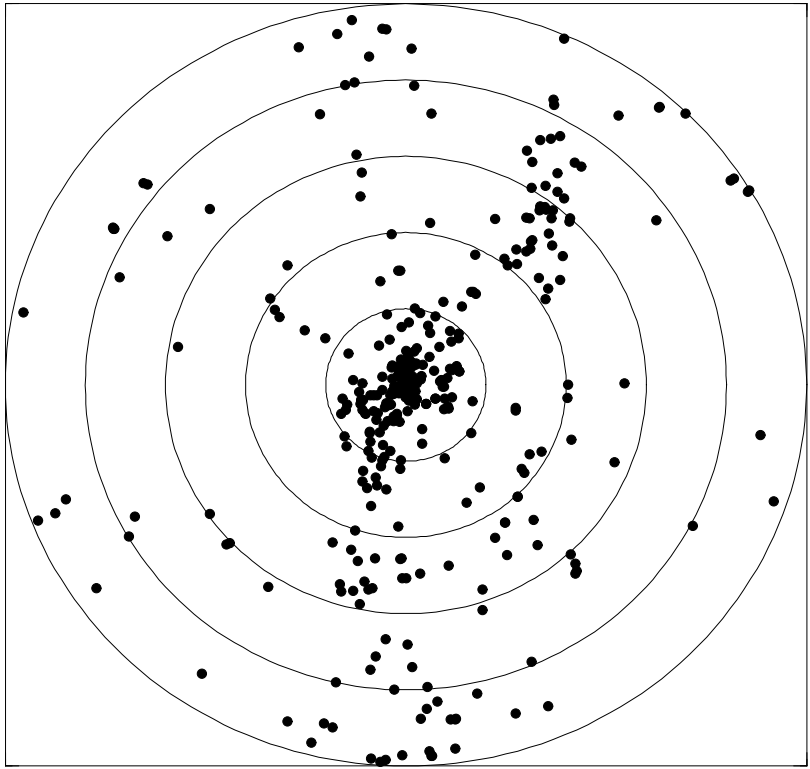,angle=270}
\end{center}
\caption{Projected view of Coma. Plotted is $(\theta\cos(\phi),
\theta\sin(\phi))$ for the same galaxies as shown the previous figure.}
\end{figure}

\begin{figure}  
\epsfig{figure=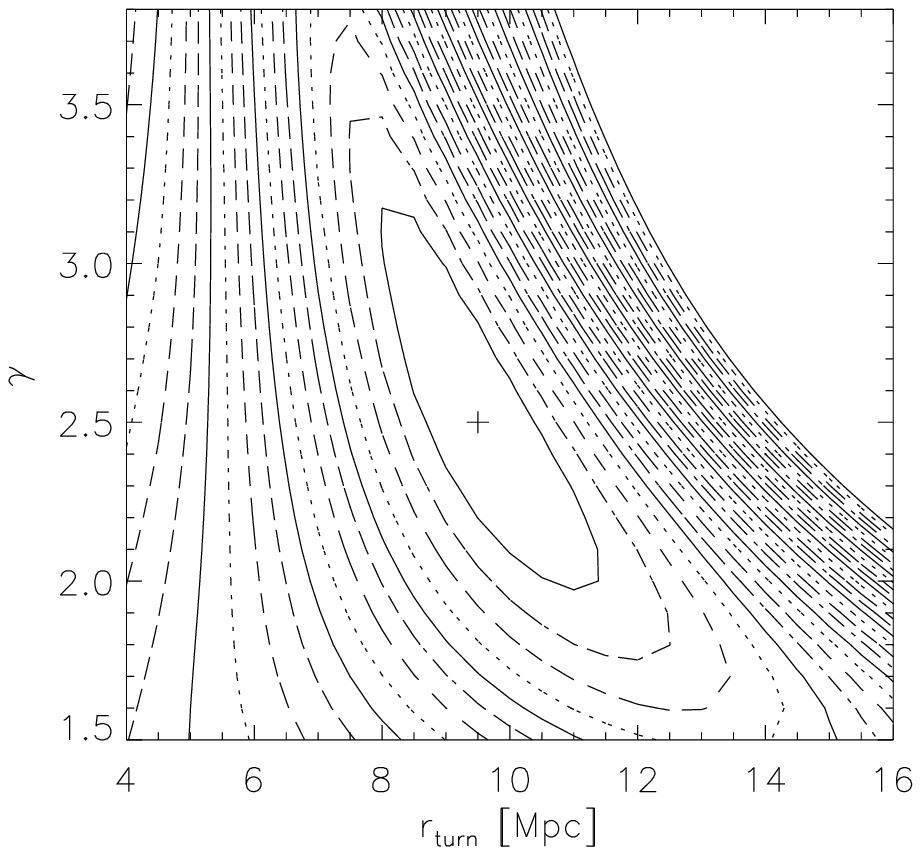,height=7.0cm}
\epsfig{figure=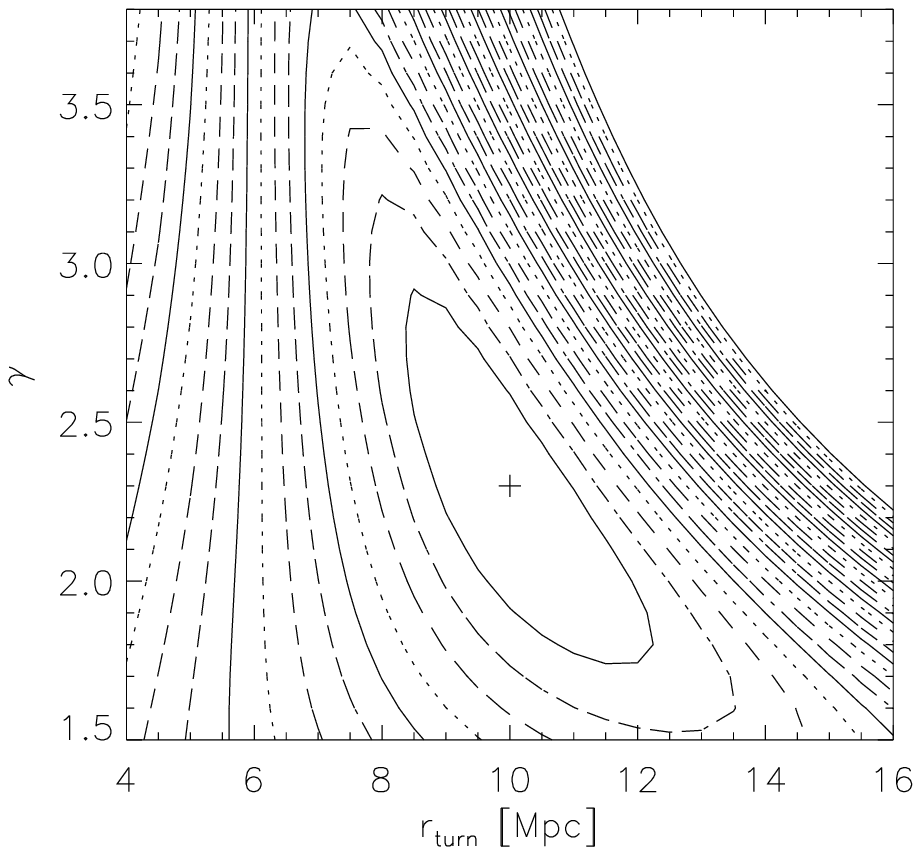,height=7.0cm}
\epsfig{figure=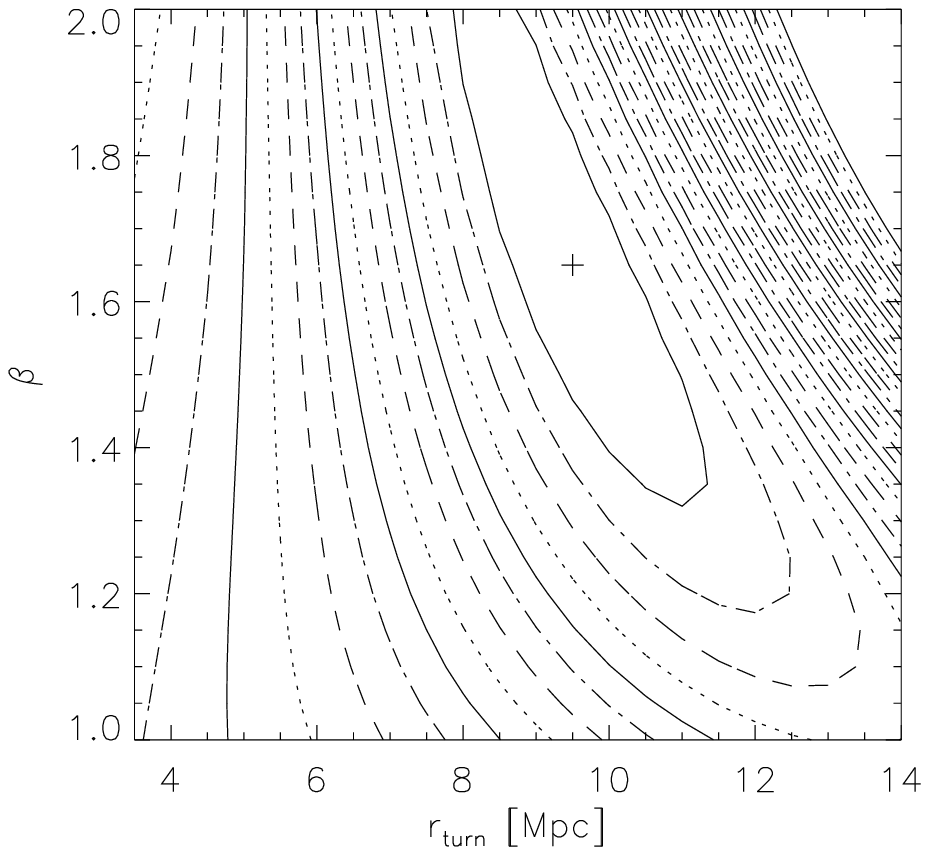,height=7.0cm}
\caption{Likelihood contours for Coma. Shown is ln(L) and contours are
separated by ln(2), maximum is indicated by cross and contours are
constantly decreasing moving away from maximum. (top left) assuming model DI,
$\sigma_{cz}=300$ km s$^{-1}$, and $\Omega_0=0.4$, (topright) DI profile and
assuming
$\sigma_{cz}=400$ km s$^{-1}$, and $\Omega_0=0.4$, (bottum) VI profile and
assuming
$\sigma_{cz}=300$ km s$^{-1}$.}
\end{figure}  

\begin{figure} 
\begin{center}
\epsfig{figure=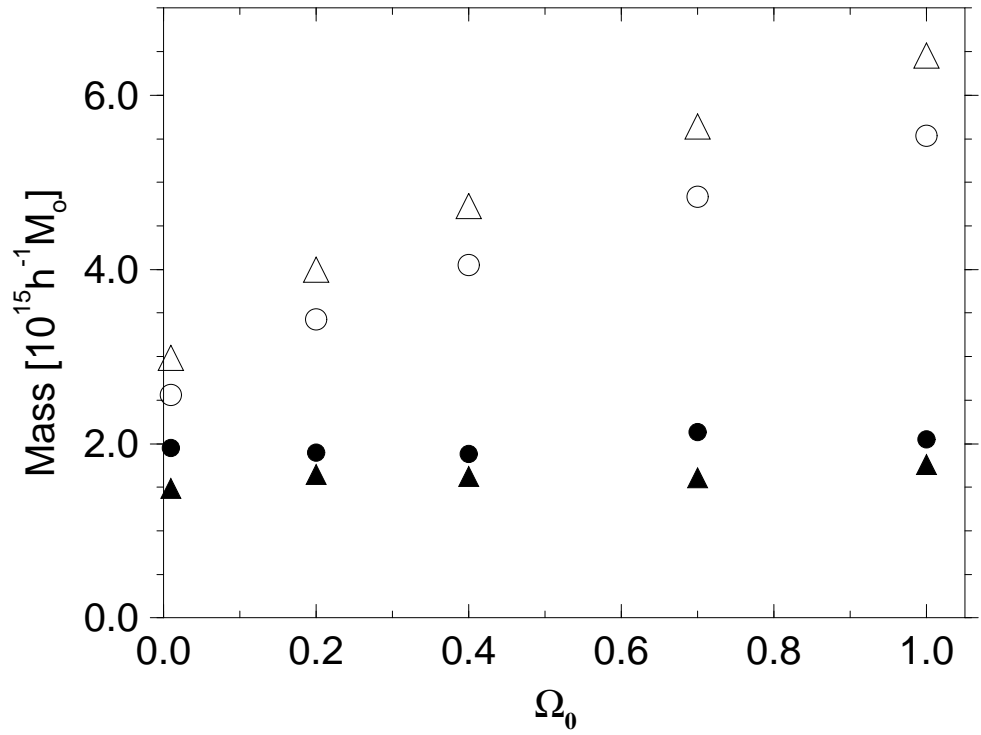,angle=270}
\end{center}
\caption{Mass estimates at $r=2.5h^{-1}Mpc$ and $r_{turn}$ of Coma based
on the DI models. Filled symbols at lower radius, circles for
$\sigma_{cz}=300$ km s$^{-1}$, triangles for $\sigma_{cz}= 400$ km
s$^{-1}$.} 
\end{figure}

\begin{figure} 
\begin{center}
\epsfig{figure=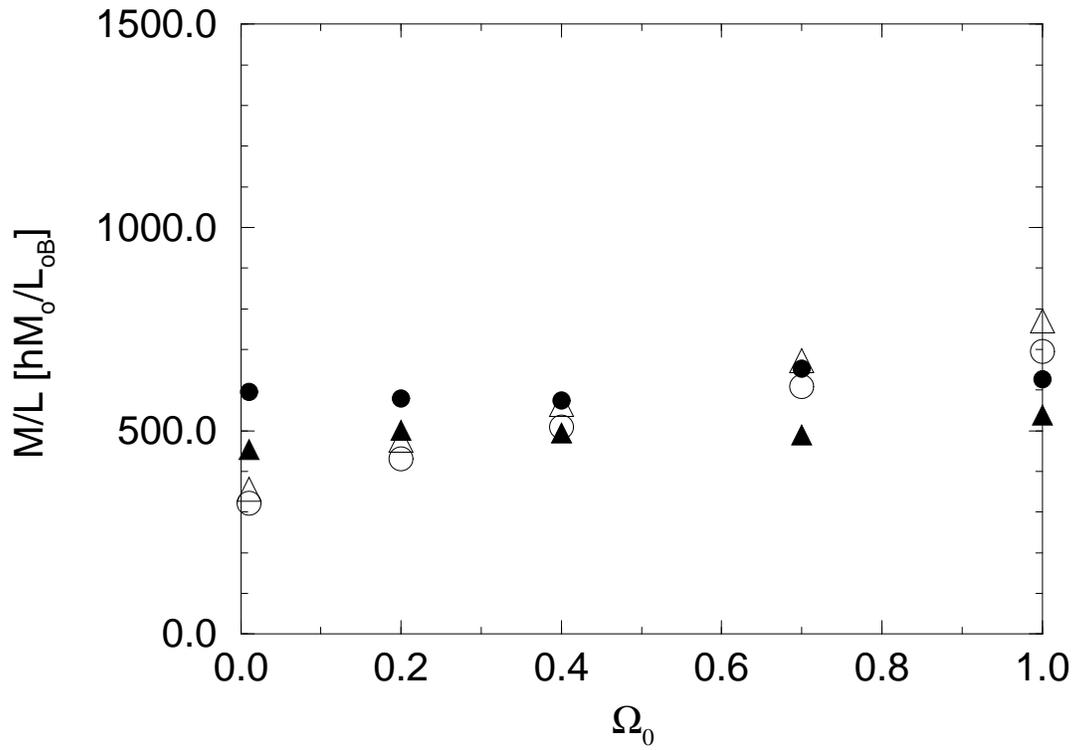,angle=270}
\end{center}
\caption{Mass-to-light ratio at $r=2.5h^{-1}Mpc$ and $r_{turn}$ for model
DI as function of $\Omega_0$ and assuming $\sigma_{cz}= 300$ km s$^{-1}$
(filled and open circles) and $\sigma_{cz}=400$ km s$^{-1}$ (filled and
open triangles).}
\end{figure}

\begin{figure} 
\begin{center}
\epsfig{figure=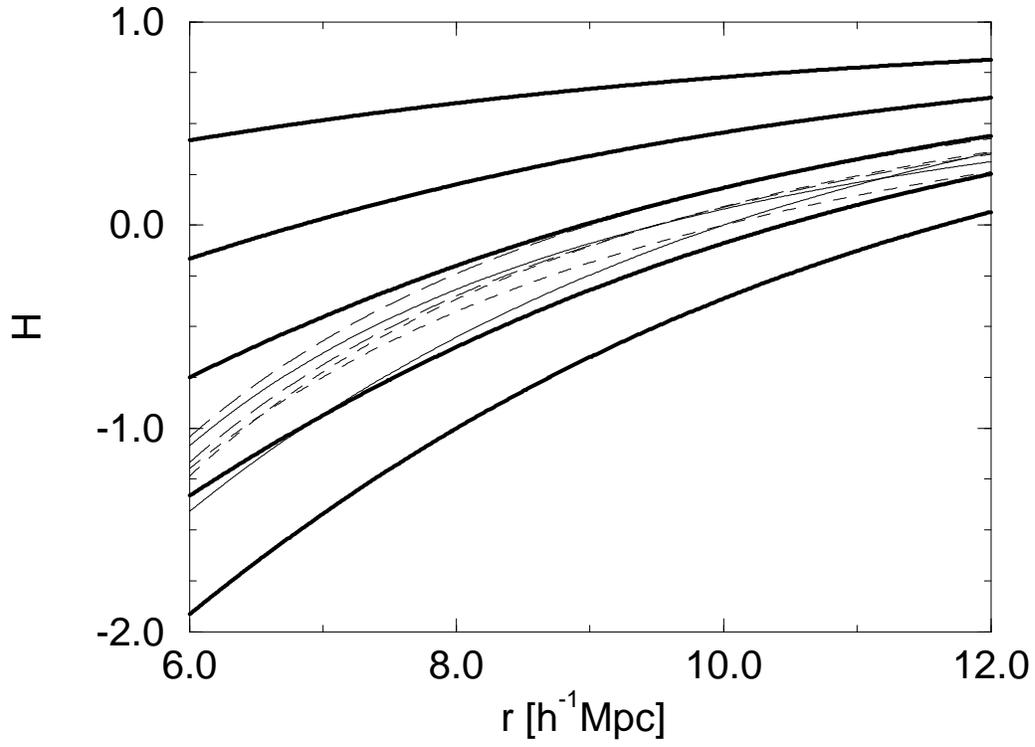,angle=270} 
\end{center}
\caption{Velocity profiles from the models (thin lines) and estimates based
on the light profile, assuming $\Omega^{0.6}/b^{0.75}$ = 1.0, 0.8, 0.6, 
0.4, and 0.2 (thick lines, starting from below). See text for
details.}
 \end{figure}

\end{document}